\documentclass{llncs}
\usepackage{amsmath}
\usepackage{amssymb}
\usepackage{graphicx}
\usepackage{graphics}
\usepackage{algorithm}
\usepackage{algorithmic}
\usepackage{amssymb}
\usepackage{url}
\usepackage{booktabs}

\urldef{\mailsa}\path|wangyi.tseg@gmail.com|\urldef{\mailsb}\path|{wubin,dunan}@bupt.edu.cn|

\begin{document}
\mainmatter  

\title{\Large Community Evolution of Social Network: Feature, Algorithm and Model
\footnotetext{This work is supported by the National Science Foundation of China under grant number 60402011, and the
National Science and Technology Support Program of China under Grant No.2006BAH03B05.}}

\author{Yi Wang \and Bin Wu \and Nan Du }

\institute{Beijing Key Laboratory of Intelligent Telecommunications Software and Multimedia, Beijing University of
Posts and Telecommunications, Beijing\\
\mailsa\\
\mailsb\\
}

\toctitle{Lecture Notes in Computer Science}

\tocauthor{Authors' Instructions}

\maketitle

\begin{abstract}
Researchers have devoted themselves to exploring static features of social networks and further discovered many
representative characteristics, such as power law in the degree distribution and assortative value used to
differentiate social networks from nonsocial ones. However, people are not satisfied with these achievements and more
and more attention has been paid on how to uncover those dynamic characteristics of social networks, especially how to
track community evolution effectively. With these interests, in the paper we firstly display some basic but dynamic
features of social networks. Then on its basis, we propose a novel core-based algorithm of tracking community
evolution, CommTracker, which depends on core nodes to establish the evolving relationships among communities at
different snapshots. With the algorithm, we discover two unique phenomena in social networks and further propose two
representative coefficients: GROWTH and METABOLISM by which we are also able to distinguish social networks from
nonsocial ones from the dynamic aspect. At last, we have developed a social network model which has the capabilities of
exhibiting two necessary features above.
\end{abstract}

\section{Introduction.}
Social network analysis has been a hot topic in the field of data mining. In the co-authorship network, a node is an
author and a edge indicates a publishing collaboration between them. Researchers are interested in these special
networks from which they discover power law in the degree distribution, that is, only a small proportion of nodes have
high degrees while the rest has low degree. Social networks also present positive assortative values while in nonsocial
networks, such as Internet, biology network, the values are always negative, indicating that in social networks, higher
degree nodes trend to connect with higher degree nodes while in nonsocial ones, it is largely possible that higher
degree ones are linked with lower degree ones. Moreover, researchers reveal community structures where the vertices
within communities have higher density of edges while vertices between communities have lower density of edges. In the
co-authorship network, a community reflects a group of scholars with similar interest. Apparently, from these static
characteristics, people have gained much understanding of social networks. However, we are not satisfied with these
achievements, but will furthermore explore those dynamic features of social networks. For example, how can we track
community evolution effectively? Does other dynamic features exist to distinguish social networks from nonsocial ones?
How can we establish a more reasonable model of social network?

With the interest to dynamic features of social networks, we firstly perform experiments in which after a long time
duration has been divided into several snapshots, we find that about 80 percent of nodes appear in one or two
snapshots. The experiment indicates that most of nodes is so unstable that we can not rely on them too much. We also
discover that node with higher degree will appear in more snapshots. On its basis, we propose a core-based algorithm
called CommTracker to track community evolution effectively. With it, we not only find out a community evolution trace
but also discover split or mergence points in the trace. By the algorithm, we find two unique phenomena of social
networks. One is that a larger community leads to a longer life and the other is that a community with a longer life
trend to have lower member stability. Correspondingly, we propose two representative coefficients: GROWTH and
METABOLISM, by which we are able to tell social networks from nonsocial ones. At last, we propose a more reasonable
model which focuses on node change. The model successfully displays two important phenomena discovered above.

We validate our conclusions in 11 datasets including 6 social networks: 3 co-authorship networks in cond-mat, math and
nonlinear fields, a call network, an email networks and a movie actor network as well as 5 nonsocial ones involving 3
software networks (tomcat 4, tomcat 5, ant), an Internet network, a vocabulary network.

The rest of the paper is organized as follows: Section 2 reviews the related work. Section 3 gives definitions. Section
4 introduces some basic dynamic features of our dataset. Section 5 presents the core-based algorithm of tracking
community evolution. Section 6 introduces two unique phenomena discovered in the social networks. Section 7 shows our
model and Section 8 concludes.

\section{Related Work.}

A lot of work has been dedicated to exploring the characteristics of social networks. Barabasi and Albert show an
uneven distribution of degree through BA models\cite{POWER_LAW}. Newman has successfully discovered distinct
characteristics between social networks and nonsocial ones\cite{assortivity}. Various methods have been utilized to
detect community structures. Among them, there are Newman's betweenness algorithm
\cite{NEWMAN_GN}\cite{NEWMAN_GN_FAST}, Nan Du's clique-based algorithm\cite{DUNAN_KDD} and CPM\cite{CPM} that focuses
on finding overlapping communities. Clustering is another technique to group similar nodes into large communities,
including L. Donetti and M.Miguel's method\cite{cluster1} which exploits spectral properties of the graph as well as
Laplacian matrix and J.~Hopcroft's ``natural community'' approach\cite{KDD03}. Some social network models have been
proposed \cite{emily_model}\cite{model2}\cite{model3}.

With respect to core node detection, Roger Guimera and  Luis A.Nunes Amaral propose a methodology that classifies nodes
into universal roles according to their pattern of intra- and inter-module connections \cite{ANOTHER_FIND_CORENODE}.
B.~Wu offers a method to detect core nodes with a threshold \cite{PEIXIN_ISDM}. Shaojie Qiao  and Qihong Liu dedicate
themselves to mining core members of a crime community\cite{911}.

As to dynamic graph mining, Tanya Y.Berger-Wolf and Jared Saia study community evolution based on node overlapping
\cite{NODE_LAP}; John Hopcroft and Omar Khan propose a method which utilizes ``nature community" to track
evolution\cite{NAR_COMM}. However, both methods have to set some parameters, which is too difficult to be adaptive to
various situations. In contrast, Keogh et al. suggests the notion of parameter free data mining\cite{NO_PARA}. Jimeng
Sun's GraphScope is a parameter-free mining method of large time-evolving graphs\cite{COMM_EVL_KDD07}, using
information theoretic principles. Our method in the paper shares the same spirit.

As forerunners, A.L.Barabasi and H.Jeong study static characteristic variations on the network of scientific
collaboration\cite{SCIE_EVOL}. Gergely Palla and A.-L. Barabasi provide a method which effectively utilizes edge
overlapping to build evolving relationship\cite{EDGE_LAP}. With the approach, they discover valuable phenomena of
social community evolution.

\section{Symbol Definition.}
The table below lists almost all the symbols used in the paper.\\
\begin{tabular}{ll}
\hline
Sym. & Definition \\
\hline
$C^{(t)}_{i}$ & Community of index $i$ in snapshot $t$\\
$N^{(t)}_{i}$ & Node of index $i$ in snapshot $t$\\
$W(N^{(t)}_{i})$ & Weight of a node of index $i$ in snapshot $t$\\
$Cen(N^{(t)}_{i})$ & Central degree of node $N^{(t)}_{i}$\\
$Core(C^{(t)}_{i})$ & Core node set of $C^{(t)}_{i}$\\
$Node(C^{(t)}_{i})$ & Node set of $C^{(t)}_{i}$\\
$Edge(C^{(t)}_{i})$ & Edge set of $C^{(t)}_{i}$\\
$|Node(C)|$ & community $C$ size\\
$C^{(t)}_{i} \to C^{(t+1)}_{j}$ &  $C^{(t)}_{i}$  is a predecessor of  $C^{(t+1)}_{j}$ or $C^{(t+1)}_{j}$ is a successor of $C^{(t)}_{i}$\\
$C^{(t-k)}_{i}\Rightarrow C^{(t)}_{j}$ & $C^{(t-k)}_{i}$ is an ancestor of $C^{(t)}_{j}$\\
$Evol(C^{(t)}_{i})$ & Evolution trace of $C^{(t)}_{i}$\\
$|Evol(C^{(t)}_{i})|$ & Span of evolution trace of $C^{(t)}_{i}$\\
\hline
\end{tabular}

\begin{definition} (COMMUNITY EVOLUTION TRACE).

An evolution trace $Evol(C^{(t)}_{x})$ is a time-series of $C^{(t+n)}$ as follows:
$$
 Evol(C^{(t)}_{x}):=\{C^{(t)}_{x},C^{(t+1)}_{x},C^{(t+1)}_{y}, C^{(t+2)}_{x} \ldots,C^{(t+n)}_{x}\} (n \geq 0)
$$
where each community $C^{(t+i)}_{x},i\in[1,n]$ satisfies the condition that there exists at least one community
$C^{(t+i-1)}_{x}$, and then $C^{(t+i-1)}_{x}\to C^{(t+i)}_{x}$. Note that more than one community is allowed to appear
in the same snapshot t+i, like $C^{(t+1)}_{x},C^{(t+1)}_{y}$ both locating in the snapshot $t+1$. $|Evol(C^{(t)}_{x})|$
is $n+1$
\end{definition}

\begin{definition} (ANCESTOR OF A COMMUNITY).

The definition of a community's ancestor is as follows: $C^{(t-k)}_{i}\Rightarrow C^{(t)}_{j}$ if there is an evolving
chain $C^{(t-k)}_{i}\to C^{(t-k+1)}_{x},\ldots,\to C^{(t)}_{j} (k \geq 1)$
\end{definition}

\begin{definition} (COMMUNITY AGE).

The age of a community is time span between its birth snapshot and its current snapshot. Here in the
$Evol(C^{(t)}_{x})$ defined in the Definition 1, the age of $C^{(t)}_{x}$ = 1 and $C^{(t+2)}_{x}$ = 3.
\end{definition}

\begin{definition} (MEMBER STABILITY OF A COMMUNITY).

The member stability of a community $C^{(t)}$ is as following:
$$
 MS(C^{(t)}) = \frac{Node(C^{(t)})\cap(Node(C^{(t+1)}_1) \cup Node(C^{(t+1)}_2) \ldots \cup Node(C^{(t+1)}_n))}{Node(C^{(t)})\cup(Node(C^{(t+1)}_1) \cup Node(C^{(t+1)}_2) \ldots \cup Node(C^{(t+1)}_n))
 }
$$
where $C^{(t)} \rightarrow C^{(t+1)}_i$ $(i\in[1,n])$
\end{definition}

\begin{definition} (MEMBER STABILITY OF A COMMUNITY EVOLUTION TRACE).

The member stability of a community evolution trace is the average stability value of all community having successors
within the trace. Its definition is as following: $\sum MS(C^{(t)})/n$, where $C^{(t)}$ is the community having
successors and $n$ is the corresponding number.

\end{definition}

\section{Basic dynamic characteristics of social networks.}

In this section, we are interested in the following three aspects: (1) how the scale of social networks evolves; (2)
how the members of social networks evolve; (3) which nodes trend to live long lives.

Note that the paper concentrates on social networks, but nonsocial networks are taken into account in that we must
compare distinct characteristics between them.

\subsection{Dataset.}

Co-authorship networks in the field of condense matter, math and nonlinear.Here, nodes represent authors and edges are
collaboration relationships of publishing papers. This three datasets include co-authorship information of Cornell
e-print from 1993 to 2006, from 1993 to 2006 and from 1994 to 2006 respectively (http://arxiv.org) and we build 28, 28
and 26 network snapshots from them by making partial dataset in half one year as a snapshot.

Cell phone network. In the network, a caller or callee is a node and the phone communication between them is an edge.
The dataset includes call information within a duration of 20 weeks in a province of China and we gain 10 network
snapshots by each including call information of 2 weeks.

Email network. Here, a node is regarded as an email sender or receiver and an edge is considered as one email
communication. This dataset from Enron (http://www.cs.cmu.edu/enron/) spans about 3 years and 32 network snapshots are
obtained, each with a duration of 1 month.

Collaboration network of movie actors. Nodes are movie actors and edges represent their collaborations. The dataset
includes collaboration information from 1980 to 2002 (http://www.imdb.com). Each snapshot is 2 years.

Internet network. From this dataset (http://sk\_aslinks.caida.org), we get 29 snapshots of Internet every 2 months.

Vocabulary network. We get vocabularies related to computer in EI Village from 1993 to 2006
(http://www.engineeringvillage2.org.cn). A node is a controlled term and if two controlled terms appear in the same
article, an edge exists between them. In this case, a snapshot lasts a year duration.

Software network of Ant, Tomcat4, Tomcat5. Here, a node represents a class and an edge exists between them if two
classes have the invoking relationship. Three datasets include 12, 19 and 21 versions respectively
(http://www.apache.org) and one version is used to establish a network.

\subsection{The evolution of network scale.}

\begin{figure*}
  \centering
  \includegraphics[width=0.9\textwidth, bb = 0 0 793 566]{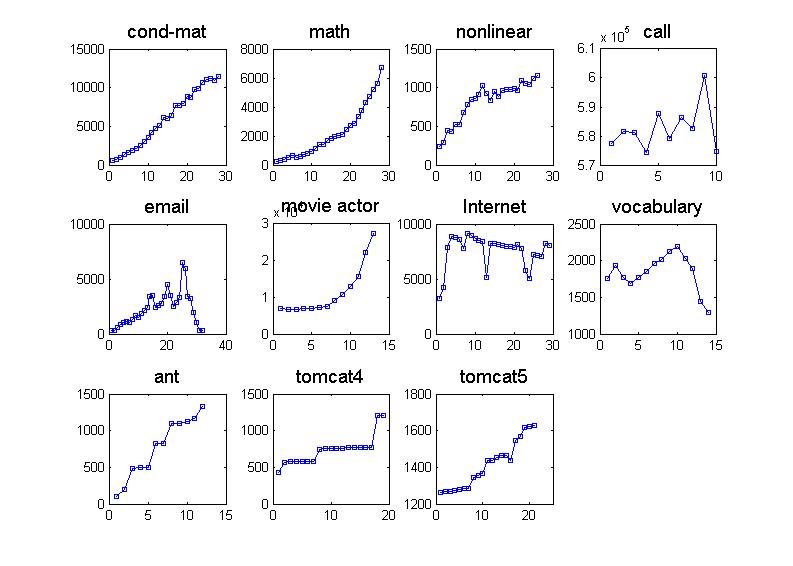}
  \caption{Network scale (node number) evolution. Snapshot id (X axis) and network scale (Y axis) }\label{network-scale}
\end{figure*}

As Fig.\ref{network-scale} shows, in each co-authorship network (cond-mat, math, non linear), the node number of
networks at different snapshot gradually increase. The phenomenon is also observed in the network of movie actor.
However, in the call network, such an increase trend is not very apparent and in the email network, we can see a
fluctuant rise, but it falls in the latest snapshots. In our analysis, co-authorship datasets and movie dataset reflect
worldwide cooperating situations, which is relatively complete. In contrast, the call network only considers the
situation of one province and the email network is from the Enron company. Both of them might reflect the partial
change. In all, we can get the conclusion that social network scale inflates when it evolves.

\subsection{The evolution of social network members.}

\begin{figure*}
  \centering
  \includegraphics[width=0.9\textwidth,bb = 0 0 793 566]{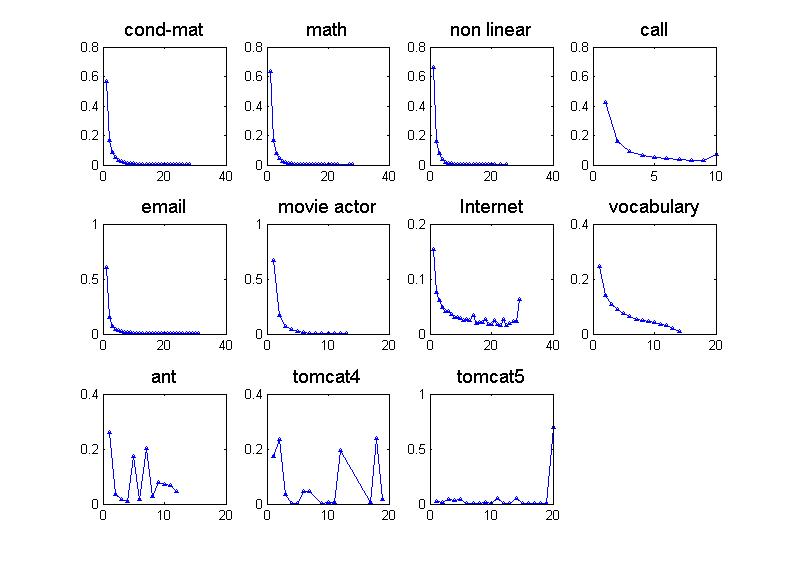}
  \caption{Node appearance distribution. Node appearance times (X axis) and percentage (Y axis)}\label{activity-percentage-correlation}
\end{figure*}

Although the size of a network increases in the evolution, its members is always changing, that is, some members will
leave the network and some will enter it. We make a statistics which indicates that during the whole evolution process,
about 80\% nodes appear in less than two snapshots (See Fig.\ref{activity-percentage-correlation}). Therefore, we
concludes that members of social networks change dramatically and only a small proportion exists in the networks
stably.

\subsection{Discovery of long life members.}

We are also interested in which nodes will get high appearance times in the network. Here, node degree is taken into
account as a critical factor, which indicates the importance of some node in the network to some extent. We
respectively calculate the correlation coefficient between node degree and appearance frequency in six social networks:
cond-mat is 0.12; math is 0.13; non linear is 0.22; call is 0.28; email is 0.44; movie actor is 0.14; In conclusion,
nodes with higher degree will exist in the network with a larger possibility.

According the conclusions in this section, we understand that a large proportion of nodes is so unstable that we can
not rely on them too much but focus on those small stable nodes, especially when we want to track community evolution.

\section{Core-based algorithm of tracking community evolution.}
As discussed above, community structures are mined by many algorithms in every network snapshots. We are interested in
how these communities evolves. For example, there exists a community in snapshot $t$, and what about its state in the
next snapshot $t+1$? Does it split into smaller ones or merge into a larger one with another community?

Our algorithm, CommTracker, heavily relies on core nodes instead of the overlapping level of nodes or edges between two
communities. From the experiments above, we have realized that most of nodes lacks stability. Therefore, taking
advantage of not all nodes that include those high fluctuating ones but these representative and reliable core nodes,
will be more accurate and effective to track community evolution. A good example is the co-authorship community where
core nodes represent famous professors and ordinary ones are other students. The research interest of professors is
usually that of a whole community. Moreover, it is harder for professors to change their research interest than for
those ordinary students.

In this section, the algorithm of core node detection is firstly introduced and then we present our core-based
algorithm of tracking community evolution.

\subsection{Core Node Detection Algorithm.}
 As discussed above, core nodes are of greatest importance in our evolution algorithm, so its preparation work,
selecting core nodes from a community, is a key step. The structure of a community is too dynamic and unpredictable to
set an empirical threshold to distinguish core nodes from ordinary ones. Unlike \cite{PEIXIN_ISDM}, the following
method concentrates on not only effectiveness but also parameter free.

A node can be weighed in terms of many aspects, such as degree, betweenness, page rank and so on. Generally, the higher
a node's weight is, the more important it is in a community. Here, we give a node $N_{i}$ a weight value $W(N_{i})$
according to its degree.

In our algorithm, both the community topology and the node weight are considered as critical factors to distinguish
core nodes from ordinary ones. In Algorithm 1, we present the whole algorithm.

\begin{figure}
  \centering
  \includegraphics[width=0.6\textwidth,bb=0 0 322 179]{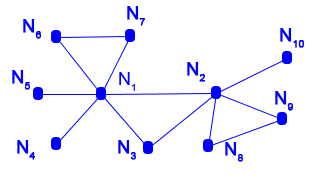}
  \caption{Core detection illustration.}\label{core_detection}
\end{figure}

The basic idea behind the algorithm is similar to a vote strategy. For each node $N_{i}$, it is entitled to evaluate
the centrality of those nodes linked with it. Assuming that $W(N_{i})$ is higher than the weight of a linked node,
$W(N_{j})$, then $N_{i}$ is considered as more important node than $N_{j}$, so $N_{i}$'s centrality value should be
incremented by a specified value while $N_{j}$'s value is reduced by a specified value. Here, $|W(N_{i})-W(N_{j})|$ is
employed to represent the centrality difference between two nodes. Through the ``vote'' of all round nodes, if
$N_{i}$'s centrality is nonnegative, it is regarded as a core node. Otherwise, it is just an ordinary node.

As Fig.\ref{core_detection} shows,  $W(N_{1})=6$. The running result is that $Cen(N_{1})=23$,\\
$Cen(N_{2})=12$ whereas $Cen(N_{4})=Cen(N_{5}) =-5$,$Cen(N_{6})= Cen(N_{7}) = Cen(N_{10})= -4$,$Cen(N_{8})= Cen(N_{9})
= -3$, $Cen(N_{3})=-7$. Therefore, the core set are $\{N_{1},N_{2}\}$.

\begin{algorithm}[!h]
\caption{CoreDetection($C$)} \label{coreDetect}
\begin{algorithmic}[1]
        \IF {$W(N_{1})=W(N_{2})=\ldots=W(N_{n})$}
            \STATE return C
        \ENDIF

    \STATE $Cen(N_{i}) = 0, i\in[1,n]$

    \FOR {every edge $e \in Edge(C)$}
        \STATE $N_{i}$,$N_{j}$ are nodes connected with $e$
        \IF {$ W(N_{i}) < W(N_{j}) $}
            \STATE$Cen(N_{i})=Cen(N_{i})-|W(N_{i})-W(N_{j})|$
            \STATE$Cen(N_{j})=Cen(N_{j})+|W(N_{i})-W(N_{j})|$
        \ENDIF
        \IF {$ W(N_{i}) \geq W(N_{j}) $}
            \STATE$Cen(N_{i})=Cen(N_{i})+|W(N_{i})-W(N_{j})|$
            \STATE$Cen(N_{j})=Cen(N_{j})-|W(N_{i})-W(N_{j})|$
        \ENDIF
    \ENDFOR

    \STATE coreset = \{\}

    \FOR{every node $N_{i}\in Node(C)$}
        \IF {$Cen(N_{i}) \geq 0$}
            \STATE input $N_{i}$ into coreset;
        \ENDIF
    \ENDFOR

    \STATE return coreset
\end{algorithmic}
\end{algorithm}

In general, Algorithm 1 is effective to detect core nodes in a small network scope, like community, where node
distances are no more than 3 hops and each node has large probability to connect to all other ones.

\subsection{Core-based Algorithm of Tracking Community Evolution.}

Tanya Y.Berger-Wolf and Jared Saia propose a method based on the overlapping level of nodes that $C^{(t+1)}$ is a
successor of $C^{(t)}$ if $nodeoverlap(C^{(t)},C^{(t+1)})\geq s$ \cite{NODE_LAP}. However, to set a proper $s$ is
challenging for users. When members of a community change dramatically and $s$ is given a higher value, $C^{(t+1)}$
will be considered to disappear because of too low overlapping level between them, but in fact $C^{(t+1)}$ still
exists. Otherwise, if $s$ is set a bit low, doing so will give irrelevant communities more opportunities to become the
successors of $C^{(t)}$, leading to ``successors explosion'' and masking those real successors.

Gergely Palla and  A.-L. Barabasi provide an approach utilizing the overlapping of edge between two
communities\cite{EDGE_LAP}, but it fails to deal with split and mergence amongst communities. As there are one
$C^{(t)}$ and two $C^{(t+1)}_i$, $C^{(t+1)}_j$, in snapshot $t$ and $t+1$ respectively, if the edge overlapping level
between $C^{(t)}$ and $C^{(t+1)}_i$ is higher than that between $C^{(t)}$ and $C^{(t+1)}_j$, $C^{(t+1)}_i$ becomes the
successor of $C^{(t)}$ while $C^{(t+1)}_j$ is considered as a new born community. Actually, $C^{(t)}$ may split into
two parts. The similar problem also exists in the process of community mergence.

The disadvantage of the method above is to treat all nodes in an unprejudiced way and it is not accorded with the
reality where different nodes have different influences. Our method has deeply paid attention to such a difference so
that it puts emphasis on core nodes.

\begin{figure}
 \centering
 \includegraphics[width=6.75cm,height=3cm,scale=0.5,bb=0 0 526 233]{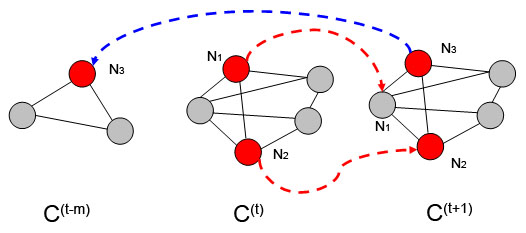}
 \caption{Community Evolution illustration: core nodes are colored red and ordinary ones grey. As we seen, (1) in snapshot $t+1$,
   $C^{(t+1)}$ contains two core node $N_1$,$N_2$ of $C^{(t)}$. (2) Node $N_3$ has also been in $C^{(t-m)}$, an ancestor of $C^{(t)}$.
   Therefore, $C^{(t+1)}$ becomes the succeeding community of $C^{(t)}$. In practice, if $C^{(t)}$ has no ancestor, then communities satisfying
  the first condition will become $C^{(t)}$'s successors automatically.}\label{example_evl}
\end{figure}

The basic thought of our algorithm can be described as:

$C^{(t)}_{i}\to C^{(t+1)}_{j}$ if and only if (1) at least one core node of $C^{(t)}_{i}$ appears in $C^{(t+1)}_{j}$,
that is, $Core(C^{(t)}_{i})\cap Node(C^{(t+1)}_{j})\neq \emptyset$ (2) at least one core node of $C^{(t+1)}_{j}$ must
appear in some ancestor community of $C^{(t)}_{i}$, that is, there exists one  $C^{(t-m)}_{k}$, $C^{(t-m)}_{k}
\Rightarrow C^{(t)}_{i}$, $Node(C^{(t-m)}_{k})\cap Core(C^{(t+1)}_{j})\neq \emptyset$. see Fig.\ref{example_evl}

\begin{algorithm}[!h]\label{coreDetect}
\caption{Community Evolution($C^{(t)}_i$)}
\begin{algorithmic}[1]
    \STATE Evol($C^{(t)}_i$) = \{$C^{(t)}_i$\}
 \STATE $Core(C^{(t)}_{i})$ = CoreDetection($C^{(t)}_{i}$)
    \FOR {every community $C^{(t+1)}_j$ in snapshot $t+1$}
        \STATE $Core(C^{(t+1)}_{j})$ = CoreDetection($C^{(t+1)}_{j}$)
        \IF {$Core(C^{(t)}_{i})\cap Node(C^{(t+1)}_{j})\neq \emptyset$ and $Node(C^{(t-m)}_{k})\cap Core(C^{(t+1)}_{j})\neq \emptyset$ and $C^{(t-m)}_{k} \Rightarrow
C^{(t)}_{i}$}
                \STATE establish the relationship $C^{(t)}_{i} \rightarrow C^{(t+1)}_j$

                \STATE Evol($C^{(t+1)}_j$) = Community Evolution($C^{(t+1)}_j$)
                \STATE Evol($C^{(t)}_i$) = Evol($C^{(t)}_i$)$\cup$Evol($C^{(t+1)}_j$)
        \ENDIF
\ENDFOR
 \STATE return $Evol(C^{(t)}_{i})$
\end{algorithmic}
\end{algorithm}

For the first condition, it is reasonable to consider $C^{(t)}_{i}$'s core nodes appear in some succeeding community
$C^{(t+1)}_{j}$, due to the representative quality of core nodes. As to the second condition, if some community
$C^{(t+1)}_{j}$ wants to become the succeeding one of a specified community $C^{(t)}_{i}$, it must suffice that its
core nodes appear in some ancestor of $C^{(t)}_{i}$, because of the stable quality of core nodes, that is , core nodes
do not appear suddenly without any evidence in the past snapshots.

We describe the whole algorithm in Algorithm 2.

From the perspective of successors and predecessors, we provide a very straightforward way to identify $community$
$split$, $community$ $mergence$, $community$ $birth$ and $community$ $death$. Note that they are four phenomena that
occurs in a single evolution trace.

\begin{itemize}
\item Community Split: a community has more than one successor.
\item Community Mergence: a community owns more than one predecessor.
\item Community Birth: a community has no predecessor.
\item Community Death: a community has no successor.
\end{itemize}

\begin{figure*}
  \centering
  \includegraphics[width=1.0\textwidth,bb=0 0 1023 158]{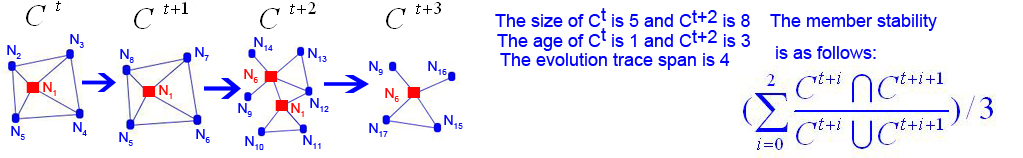}
  \caption{Community evolution illustration. Red square points are core nodes. }\label{community_evolution_illustration}
\end{figure*}

Fig.\ref{community_evolution_illustration} shows a typical example of community evolution.

\section{Two representative phenomena in the social network.}

\begin{figure*}
  \centering
  \includegraphics[width=1.0\textwidth,bb=0 0 800 573]{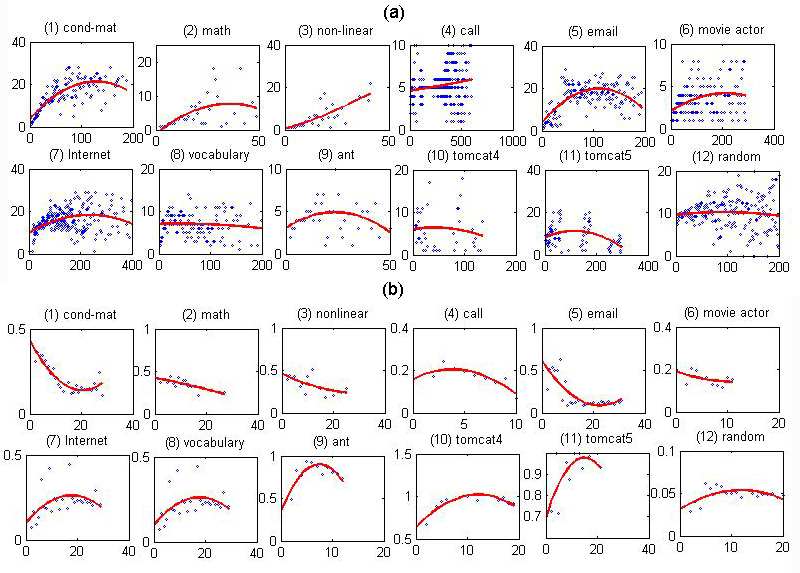}
  \caption{(a) The correlation between community size (X axis) and community age (Y axis).
   (b) The correlation between evolution trace span (X axis) and member stability (Y axis). }\label{2_phenomena}
\end{figure*}

In \cite{EDGE_LAP}, Palla has performed two experiments only on cond-mat co-authorship and call networks: one is to
find out the correlation between community size and age; the other is to uncover the correlation between evolution
trace span and member stability. In his paper, he obtains conclusions that communities of larger size lead to longer
lives and that if an evolution trace span is longer, its member stability is lower. We are interested in the two
situations in other social networks and nonsocial ones. The results are shown in Fig.\ref{2_phenomena} (a) and (b).

Firstly, depending on CommTracker, we can discover similar phenomena with those proposed in the Palla's paper, proving
that our method is effective and correct. Secondly, it is obvious that 6 social networks display two common behaviors
we discuss above. On the contrary, nonsocial networks fails to own such behaviors. In nonsocial networks, it seems that
the size of a community can not reflect its age and that a community with higher stability will live for a longer life.

We calculate the correlation coefficients between community size and age (GROWTH) as well as between evolution trace
span and member stability (METABOLISM) (See Table \ref{growth_metabolism}). Apparently, in the 1st experiment, social
networks' values are positive while those of nonsocial ones are nearly all negative. In the 2nd experiment, the values
of social networks are negative whereas those of nonsocial ones are all positive. Two experiments reveal that we can
differentiate social networks from nonsocial ones according to GROWTH and METABOLISM.

\begin{table}
  \centering
\begin{tabular}{|c|c|c|c|c|c|c|c|c|c|c|c|c|}
    \hline
    &1&2&3&4&5&6&7&8&9&10&11&12\\
    \hline
    GROWTH & 0.67 & 0.45 & 0.76 & 0.2 & 0.39 & 0.31 & 0.29 & -0.07 &-0.02 & -0.09 & -0.23 & -0.01\\
    \hline
    METABOLISM & -0.76 & -0.72 & -0.62 & -0.76 & -0.67 & -0.37 & 0.25 & 0.25 & 0.23 & 0.47 & 0.51 & 0.16\\
    \hline
\end{tabular}
\caption{GROWTH and METABOLISM. (1) cond-mat (2) math (3) nonlinear (4) call (5) email (6) movie actor (7) Internet (8)
vocabulary (9) ant (10) tomcat4 (11) tomcat5 (12) random }\label{growth_metabolism}
\end{table}

One important reason contributing to such distinctions is that in social networks, a community represents a group of
persons with close connection and in nonsocial ones a community is just a cluster of objects. As we know, in social
networks, if a community want to obtain a long life, it must undertake suitable member changes, that is, when some old
core members retire, new ones take over responsibility in time so that the development of the community is well
supported. Otherwise, if a community refuses to absorb new members, when the old core members exit from the community,
it is possible that new core ones have not been cultivated, leading to quick disintegration. In contrast, the members
of nonsocial networks are objects, not persons. For example, in software network, a community is a class cluster with
similar functions. If a class cluster is designed well, it must experience little change and be used for a long time.

\section{Social network model.}
Nowadays, many social network models have been established.  However, when we get some snapshots generated from these
social networks, most of them fail to display the characteristic behaviors we have proposed above. In our view, a main
defect is that a node will permanently exist in the network once it is added into the network. However, from the
experiments shown in Section 4, a lot of nodes enter into the network and then quickly exist from it. Hence, how to
revise existing models to make them more reasonable is a problem to be solved.

\subsection{Model introduction.}
Our model is based on the one proposed in Emily's model\cite{emily_model}, which takes into account social network
aspects completely, such as meeting rate between pairs of individuals, decay of friendships, etc. Moreover, Emily's
model indeed presents many static features of social networks. Therefore, we decide to adopt it as our model basis. Our
model can be simulated directly using the following algorithm.

Let $n_p=\frac{1}{2}N(N-1)$ where $N$ is the network initial scale. Let $n_e=\frac{1}{2} \sum z_i$ where $z_i$ is the
degree of the $i^{th}$ vertex. And let $n_m=\frac{1}{2} \sum z_i(z_i-1)$.

1. We choose $n_p r_0$ pairs of vertices uniformly at random from the network to meet. If a pair meet who do not have a
pre-existing connection, and if neither of them already has the maximum $z^*$ connections then a new connection is
established between them.

2. We choose $n_m r_1$ vertices at random, with probability proportional to $z_i(z_i-1)$. For each vertex chosen we
randomly choose one pair of its neighbors to meet, and establish a new connection between them if they do not have a
pre-existing connection and if neither of them already has the maximum number $z^*$ of connections.

3. We choose $n_e \gamma$ vertices with probability proportional to $z_i$. For each vertex chosen we choose one of its
neighbors uniformly at random and delete the connection to that neighbor.

4. We choose one vertex, if its degree $z_i > \overline{z}$, the average degree, we delete it with the probability
$\alpha$; otherwise, we delete it with the probability $\beta$. The process doesn't stop until $k_d$ vertices have been
deleted.

5. We add $k_a$ new vertices. For each new one, it establishes a link with a vertex $v$ randomly and then it also
connects to the vertex with highest degree from the neighbor vertices of $v$.

Note that the first 3 steps have already existed in the Emily's algorithm while the last 2 steps are added by
ourselves. The 4th step is responsible for deleting some existing vertices according to their degrees. The last step
focuses on adding new vertices. In this step, we eliminate the limit of maximum connection in order to allow some
vertices to get high degree. In reality, a community consists of vertices with distinct degrees while in the Emily's
social network model, a community trends to be a clique due to the limit of maximum connection.

As pointed out in \cite{emily_model}, the network is initialized by starting with no edges, and running the first two
steps without the other three ones until all or most vertices have degree $z^*$ (we set the limitation as 85\%). Then
all five steps are used for the remainder of the simulation.

\subsection{Model stimulation.}

\begin{figure*}
  \centering
  \includegraphics[width=1.0\textwidth,bb = 0 0 793 566]{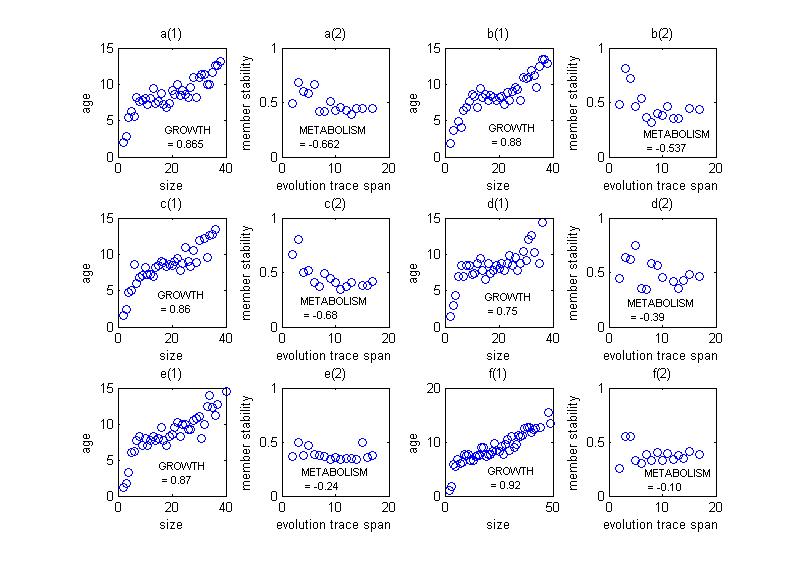}
  \caption{Model stimulation. (a) $\alpha = 0.8$,$\beta = 0.6$,$k_a = k_d = 3$ (b)  $\alpha = 0.5$,$\beta = 0.5$,$k_a = k_d = 3$
  (c) $\alpha = 0.3$,$\beta = 0.8$,$k_a = k_d = 3$ (d) $\alpha = 0.8$,$\beta = 0.3$,$k_a = k_d = 3$ (e) $\alpha = 0.5$,$\beta = 0.5$,$k_a = k_d = 6$
  (f) $\alpha = 0.5$,$\beta = 0.5$,$k_a = 5$, $k_d = 3$  }\label{model}
\end{figure*}

Six experiments have been performed with different parameters $\alpha$, $\beta$, $k_a$ and  $k_d$ shown in
Fig.\ref{model}. In all stimulation, $z^*=5$, $N=250$, $r_0=0.0005$, $r_1=2$, $\gamma=0.005$. When all the five steps
are running, we get a snapshot every five repetitions. We consider 17 snapshots together.

\section{Conclusions.}

In the paper, we firstly perform some basic experiments to explore those dynamic characteristics of social networks and
it is discovered that a large percentage of nodes are so instable that we can not rest on them too much and that nodes
with higher degree will appear more frequently during the evolution of a social network. Under the experimental
results, we propose a novel core-based algorithm to track community evolution, which has the following features:(1) it
is effective; (1) it is parameter-free; (2) it is suitable to discover split and mergence points. With the algorithm,
we uncover two representative dynamic features of social networks and define two coefficients: GROWTH and METABOLISM by
which we also achieve the goal of telling social networks from nonsocial ones. In the end, we propose a revised social
network model which can display two typical characteristics. The experiments are based on 6 social networks
(co-authorship network, call network, movie actor network and email network)and 5 nonsocial networks (Internet,
vocabulary network and software network).

{}


\begin{thebibliography}{}
\bibitem{POWER_LAW}
A.~L. Barabasi and R.~Albert, {\em Emergence of scaling in random networks}, Science, 1999, pp.~509--512

\bibitem{CLUSTER_DEGREE}
D.~J. Watts and S.~H.Strogatz, {\em Collective dynamics of small world network}, Nature, 1998, pp.~440--442

\bibitem{PEIXIN_ISDM}
B.~Wu, X.~Pei, J.~Tan, and Y.~Wang, {\em Resume Mining of Communities in Social Network}, In ICDM Workshop, 2007

\bibitem{ANOTHER_FIND_CORENODE}
R.~Guimera, L.~A. Nunes Amaral, {\em Functional cartography of complex metabolic networks}, Nature J., 2(2005),
pp.~895--900

\bibitem{cluster1}
L.~Donetti and M.~Miguel,{\em Detecting network communities: a new systematic and efficient algorithm}, Journal of
Statistical Mechanics, 2004, pp.~100--102

\bibitem{NODE_LAP}
Tanya Y.Berger-Wolf and Jared Saia, {\em A framwork for analysis of dynamic social network}, In KDD, 2006

\bibitem{EDGE_LAP}
G.~Palla and A.~L. Barabasi, {\em Quantitying social group evolution}, Nature J., 5(2007),
  pp.~664--667.

\bibitem{COMM_EVL_KDD07}
J.~Sun, P.~S. Yu, S.~Papadimitriou, and C.~Faloutsos, {\em
  GraphScope:Parameter-free Mining of Large Time-evolving Graphs},
  In KDD, 2007, pp.~687--696.

\bibitem{NAR_COMM}
J.~Hopcroft, O.~Khan, B.~Kulis, B.~Selman, {\em
  Tracking evolving communities in large linked networks},
  PNAS, 2004, pp.~5249-5253

\bibitem{KDD03}
J.~Hopcroft, O.~Khan, B.~Kulis, B.~Selman, {\em Natural Community in Large Linked Networks}, In KDD, 2003

\bibitem{CPM}
J.~Palla, G.~Vicesek, T.~Vicsek, B.~Selman, {\em
  Uncovering the overlapping community structure of complex network in nature and society},
  Nature J., 2005,
  pp.~814--818.

\bibitem{DUNAN_KDD}
N.~Du, B.~Wu, X.~Pei, B.~Wang, L.~Xu {\em
  Community Detection in Large-Scale Social Network},
  In KDD workshop, 2007

\bibitem{NEWMAN_GN}
M.~Givan, M.~E.J. Newman, {\em Community structure in social and biological networks}, PNAS, 2002, pp.~7821--7826

\bibitem{NEWMAN_GN_FAST}
M.~E.J. Newman, {\em Fast algorithm for detecting community structure in networks}, Phys.Rev.E69, 066133(2004)


\bibitem{NO_PARA}
E.~Keogh, S.~Lonardi, C.~A. Ratanamahatana. {\em
    Towards parameter-free data mining},
    In KDD, 2004, pp.~206--215.

\bibitem{COMM_GROWTH}
L.~Backstrom, D.~P.Huttenlocher, J.~M.Kleinberg, X.~Lan {\em Group formation in large social networks: membership,
growth, and evolution.}, In KDD, 2006, pp.~44--54.

\bibitem{COMM_SHRINK}
J.~Leskovec, J.~Kleinberg and C.~Faloutsos. {\em Graphs over time:Densification laws, shrinking diameters and possible
explanations.}, In KDD, 2005,

\bibitem{SCIE_EVOL}
A.~L.Barabasi, H.~Jeong, Z.~Neda, E.~Ravasz, A.~Schubert, T.~Vicsek. {\em Evolution of the social network of scientific
collaborations}, arXiv:cond-mat/0104162v1

\bibitem{911}
Q.~Liu, C.~Tang, S.~Qiao, Q.~Liu, F.~Wen. {\em Mining the Core Member of Terrorist Crime Group Based on Social Network
Analysis}, PAISI, 2007, pp.~311--313

\bibitem{assortivity}
M.~E.J. Newman, {\em Assortative mixing in networks}, Physical Review Letters, 89(20):208701.

\bibitem{emily_model}
E.~M. Jin, M.~Girvan, M.~E.J.Newman(2001). {\em The structure of growing social networks}, Physical Review Letters, 64,
046132

\bibitem{model2}
A.~ Gronlund, P.~Holme, {\em The networked seceder model: Group formation in social and economic systems}, Phys. Rev. E
70, 036108 (2004)

\bibitem{model3}
B.~Skyrms, R.~Pemantle, {\em A Dynamic Model of Social Network Formation}, PNAS, 96 (16), pp.~9340-9346

\end{thebibliography}
\end{document}